# From fractional Chern insulators to topological electronic crystals in moiré MoTe2: quantum geometry tuning via remote layer

Feng Liu[1†], Fan Xu[1,2†], Cheng Xu[3†], Jiayi Li[1], Zheng Sun[1], Jiayong Xiao[1], Ning Mao[3], Xumin Chang[1], Xinglin Tao[1], Kenji Watanabe[4], Takashi Taniguchi[5], Jinfeng Jia[1,2,6,7], Ruidan Zhong[1,2], Zhiwen Shi[1,2], Shiyong Wang[1,2], Guorui Chen[1,2], Xiaoxue Liu,[1,2,6] Dong Qian[1,2], Yang Zhang[8,9*], Tingxin Li[1,2,6*] & Shengwei Jiang[1,2*]

[1]Key Laboratory of Artificial Structures and Quantum Control (Ministry of Education), School of Physics and Astronomy, Shanghai Jiao Tong University, Shanghai 200240, China

[2]Tsung-Dao Lee Institute, Shanghai Jiao Tong University, Shanghai, 201210, China

[3]Max Planck Institute for Chemical Physics of Solids, 01187, Dresden, Germany

[4]Research Center for Electronic and Optical Materials, National Institute for Materials Science, 1-1 Namiki, Tsukuba 305-0044, Japan

[5]Research Center for Materials Nanoarchitectonics, National Institute for Materials Science,  1-1 Namiki, Tsukuba 305-0044, Japan

[6]Hefei National Laboratory, Hefei 230088, China

[7]Shanghai Research Center for Quantum Sciences, Shanghai 201315, China

[8]Department of Physics and Astronomy, University of Tennessee, Knoxville, TN 37996, USA

[9]Min H. Kao Department of Electrical Engineering and Computer Science, University of Tennessee, Knoxville, Tennessee 37996, USA

[†]These authors contribute equally to this work.
[*]Emails: yangzhang@utk.edu, txli89@sjtu.edu.cn, swjiang@sjtu.edu.cn

**Abstract**
The quantum geometry of Bloch wavefunctions, encoded in the Berry curvature and quantum metric, is believed to be a decisive ingredient in stabilizing fractional quantum anomalous Hall (FQAH) effect (i.e., fractional Chern insulator, FCI, at zero magnetic field)[1–11], against competing symmetry-breaking phases[12–22]. A direct experimental demonstration of quantum geometry-driven switching between distinct correlated topological phases, however, has been lacking. Here, we report experimental evidence of such a switch in a high-quality 3.7° twisted MoTe2 (*t*MoTe2) device consisting of both A-A bilayer and A-AB trilayer regions. While composite Fermi liquid CFL[23]/FQAH phases are established in A-A *t*MoTe2, the A-AB region—effectively an A-A moiré bilayer proximitized by a remote B layer—develops a series of topological electronic crystal (TEC, also referred to as generalized QAH crystal, QAHC) states with integer quantized Hall conductance at commensurate fractional fillings $\nu_h = 1/2$, 2/3, and an incommensurate filling factor $\nu_h \approx 0.53$. The electrostatic phase diagram is mapped out by combined transport and optical measurements, showing that these

TEC states emerge within the first moiré valence band prior to any charge transfer to the B layer. Exact diagonalization (ED) incorporating the remote-layer-induced intralayer potential demonstrates a transition from a CFL–like manifold in the A-A limit to a Chern number $C = 1$ ground-state consistent with a TEC at $\nu_h = 1/2$, accompanied by the further breakdown of ideal band geometry. Our results provide experimental evidence of quantum geometry-tuned competition between FQAH/CFL and TEC phases in a moiré Chern band and pave the way for further exploring correlation-driven topological phenomena by tuning quantum geometry.

**Main**

The behavior of electrons in quantum materials depends sensitively on the underlying quantum geometry $\boldsymbol{T} = \boldsymbol{g} - i/2\boldsymbol{\Omega}$—namely the distribution of Berry curvature $\boldsymbol{\Omega}$ and the quantum metric $\boldsymbol{g}$—of the constituent Bloch states[24]. The quantum geometry is believed to play a critical role in the realization of the FQAH phases against other competing correlated phases[12–16,19–22]. In general, FCI in an electronic band can be realized if the band is topologically nontrivial and the dispersion is flat enough, such that correlations may stabilize charge gaps at fractional fillings[25]. However, these conditions are insufficient to guarantee the emergence of FCI. At commensurate fractional fillings of a topological flat band, a topological charge gap may arise in two distinct ways (we acknowledge there is another way proposed recently, i.e., fractional QAHCs)[26]: (i) an FCI supporting fractionally charged excitations and a fractional many-body Chern number $C = \nu C_{\text{band}}$, where $C_{\text{band}}$ is the Chern number of the parent band; or (ii) a symmetry-breaking electronic crystal that enlarges the real-space unit cell so that an integer number of carriers binds to the emergent supercell, enforcing an integer $C$ despite fractional filling of the original moiré cell. These scenarios generally compete, and near-ideal quantum geometry (e.g., satisfying the trace condition $\text{Tr}\boldsymbol{g} = |\boldsymbol{\Omega}|$,) favors the former[12–14,19–21,27]. Instead of the requisite breaking continuous translation symmetry in the original Hall crystal in the two-dimensional electron gas (2DEG)[28], a generalized version, TEC (or generalized QAHC), could arise more readily in moiré systems, exhibiting both a charge gap and zero-field quantized AH at fractional fillings[29–32].

Recent theory predicts that $t$MoTe$_2$ can host both FQAH and TEC phases, with the outcome sensitive to quantum geometric conditions[17,18]. Yet, experimental observation of a quantum geometry-driven phase switch between these competing topological orders has remained elusive. In this study, we report a combined transport and optical study of a high-quality 3.7° $t$MoTe$_2$ device consisting of both A-A and A-AB regions. By introducing a remote B MoTe$_2$ layer, we observed the ground state switches from CFL/FQAH to TEC with integer quantized Hall conductance at both commensurate and incommensurate filling factors. We mapped out the electrostatic phase diagram across different layer polarization regimes, confirming that the switching from CFL/FQAH to TEC, observed between A-A and A-AB $t$MoTe$_2$, occurs within the first moiré band derived from the A-A bilayer.

**Basic characterization of A-AB in comparison with A-A**

In this work, we study twisted monolayer–bilayer MoTe$_2$, where monolayer and bilayer MoTe$_2$ are stacked and rotated by a small angle with respect to each other (see Extended Data Fig.1). Fig. 1a shows the schematic of our dual-gated device consisting of both A-A twisted bilayer MoTe$_2$ region and A-AB twisted trilayer region with the same twist angle, fabricated following the similar procedure reported earlier[4,33] (see Methods). The doping density and twist angle of 3.71° (±0.08°) are independently calibrated by the quantum oscillation observed under high magnetic fields (Extended Data Fig. 2). The applied electric field ($E$) was determined from the gate voltages using the parallel-plate capacitor model (see Methods).

Schematic illustrations of the band structures (valence band only) for A-A and A-AB are shown in Fig. 1b and 1c. In the vicinity of the K point of the Brillouin zone, the band structure of each MoTe$_2$ layer can be regarded as a single spin-polarized parabolic band. Electron hopping between the AA stacked two layers caused a strong hybridization of the bands where they intersect, leading to a low-energy moiré miniband. In the A-AB region, the hybridization is much stronger between the A-A moiré bilayer than the A-B natural bilayer, due to the constraint of spin-valley conservation of interlayer hopping. Therefore, A-AB is effectively an A-A moiré bilayer plus a weakly coupled B layer. The strong hybridization of the A-A bilayer also leads to a smaller band gap between conduction and valence bands compared with its monolayer counterparts, as reported in $t$MoTe$_2$ and $t$WSe$_2$[34,35]. Therefore, the higher-energy band from the B layer acts as a remote band for the first moiré band (Fig. 1c). This band alignment is supported by our density functional theory (DFT) calculations on A-AB $t$MoTe$_2$ (Fig. 1d) as well as the experimental results discussed below, and consistent with recent theoretical calculation[36]. As a result, at small $E$ for $v_h \leq 1$, all the holes will only occupy the first moiré band. We also note that the band dispersion is substantially modified in A-AB, showing momentum-dependent spin splitting, stemming from the B-layer-induced broken $C_{2y}$ symmetry.

To investigate the influence of the B layer, we first performed reflective magnetic circular dichroism (MCD) measurements under a small out-of-plane magnetic field ($B$) on both the A-A and A-AB regions (see the Methods).

Fig. 1e shows the MCD intensity map plot as a function of $v_h$ and $E$ of the A-A region. Strong MCD signal suggests the continuous ferromagnetic phase over a wide range of filling factor of ~0.4 to 1.2. The MCD versus $B$ curves show pronounced hysteresis loops at $v_h = 1$, $2/3$, and $1/2$ (Extended Data Fig. 3), a hallmark of ferromagnetism. Notably, the critical electric field $E_c$ for suppressing the FM state is enhanced at the fractional filling $v_h = 2/3$. The FCI nature of $v_h = 2/3$ can be examined by measuring the photoluminescence (PL) intensity as a function of $v_h$ and B. The dispersion of the $v_h = 1$ and $2/3$ states agrees with expected Chern number $C = 1$ and $2/3$, according to the Streda formula $n_M \frac{dv_h}{dB} = C \frac{e}{h}$ (Extended Data Fig. 3). where $n_M$ is the moiré density, $e$ is the elementary charge, and $h$ is Planck's constant. The above results are fully consistent with the previous report[1–4].

Remarkably, the A-AB region shows substantially different magnetic response, as shown in Fig. 1f. Firstly, the MCD response is highly asymmetric with respect to $E$, which is expected considering the broken $C_{2y}$ symmetry of the A-AB lattice structure. Secondly, the phase spacing showing strong MCD response becomes discontinuous: the MCD response is strongly suppressed around $\nu_h = 2/3$ (Extended Data Fig. 4), while a local maximum is observed around $\nu_h = 1/2$. Since A-A and A-AB regions share the same piece of A-A moiré bilayer, the sharp contrast in their magnetic response reflects that the low-energy physics in the first moiré band is significantly affected by the nearby B layer.

**Electrostatic phase diagram**

We performed magneto-transport and optical spectroscopy measurements to determine the electrostatic phase diagram and identify the charge distribution at filling factors of interest. Figures 2a and 2b show the longitudinal resistance, $R_{xx}$, as a function of $(\nu_h, E)$ under out-of-plane magnetic field $B = 0.1\,\text{T}$ and $10\,\text{T}$, respectively, at temperature $T = 1.6\,\text{K}$. Figure 2c shows the schematic electrostatic phase diagram extracted from the combined transport and optical spectroscopy measurements. In figure 2a, at $\nu_h = 1, 2/3, 1/2, 1/3$ and $1/4$, local peaks in $R_{xx}$ can be identified at positive $E$-field, corresponding to the topologically trivial correlated insulator states. The $\nu_h = 1/3$ and $1/4$ correlated insulating states have rarely been previously reported in $t$MoTe2, together with the high mobility (Extended Data Fig. 2), highlighting the high quality of our device. Another pronounced feature is a continuous boundary of $R_{xx}$, across which $R_{xx}$ drops from the relatively positive $E$-field side to the negative $E$-field side. This boundary (denoted by the red solid line in Fig. 2c) corresponds to the threshold of $E$-field ($E_c$) above which the holes start to transfer from the A-A bilayer to the B layer. The above interpretation is also supported by the magneto-transport measurement. In Figure 2b, quantum oscillations (stripes) could be observed in many regions of the phase diagram. For $\nu_h < 1$, we identify a continuous region with pronounced vertical stripes (region A1+A2 and region II). In these regions, the Landau levels do not disperse with $E$-field, that is, $E$-field does not induce charge transfer between adjacent layers. By contrast, in regions with larger $E$-field beneath region II, i.e., region A2+B and region A1+A2+B, the vertical stripes turn into diagonal ones, signifying the charge transfer between the A-A and the B. The above observation can be understood by the band alignment of the A-AB $t$MoTe2, i.e., the first moiré band has lower energy than the B-layer band due to the stronger hybridization, resulting in a charge transfer gap (Fig. 1d). The application of negative $E$-field reduces and eventually closes the gap. When the Fermi level cuts through both bands, the charge transfer happens. The charge transfer gap can be estimated by $\Delta_C(\nu_h) = D_e \cdot E_c(\nu_h)$, where $D_e$ is the effective interlayer electric dipole moment is $D_e \approx 0.16\,\text{e}\cdot\text{nm}$ is extracted from the optical measurements (Extended Data Fig. 5), in good agreement with the estimated $D \approx 0.17\,\text{e}\cdot\text{nm}$ from electrostatics (see Methods). Therefore, we estimate $\Delta_C\left(\frac{1}{2}\right) \approx 8\,\text{meV}$, $\Delta_C\left(\frac{2}{3}\right) \approx 6\,\text{meV}$ and $\Delta_C(1) \approx 3\,\text{meV}$.

When $B$-field is reduced to 5 T, the quantum oscillations vanish in region A1+A2 and region A2 but are still prominent in region A2+B and region A1+A2+B (Extended Data Fig. 6), signifying the heavier effective mass of the first moiré band compared with the B-layer band.

Figures 2d and 2e show the RC spectra as a function of $E$-fields at $v_h = 1$ and $1/2$ respectively. The corresponding layer polarization states are divided by the horizontal dashed lines. At large positive $E$, all the holes are polarized into layer A1 (also denoted as region A1 in Fig. 2c), we would expect to see the neutral excitons of layer A2 ($X_{A2}$) and B ($X_B$). At large negative $E$, all the holes are polarized into layer B, we would expect to see the neutral excitons of layer A1 ($X_{A1}$) and $X_{A2}$. Bearing the above discussion in mind, the layer correspondence of the exciton resonances at large $E$ can be readily identified, as denoted in the figures. At both $v_h = 1$ and $1/2$, $X_B$ shows the highest resonance energy, consistent with the B layer having a larger band gap than the A+A moiré bilayer. $X_{A2}$ and $X_{A1}$ show different resonance energy, likely due to their different dielectric environment (A2 in the middle while A1 on the bottom). Based on the extracted electrostatic phase diagram, holes only occupy the first moiré band in region A1+A2, i.e., the B-layer band should be empty, and layer B should be neutral.

**QAH states at commensurate and incommensurate fractional fillings**

Inspired by the MCD results, we further performed transport measurements at millikelvin temperature to examine the ground states of the A-AB system. Here we focus on the first moiré band, i.e., region A1+A2, where holes reside only in the A-A moiré bilayer.

Figures 3a and 3b show longitudinal resistance $R_{xx}$ and Hall Resistance $R_{xy}$ as a function of ($v_h, E$) under an out-of-plane $B = 1$ T, respectively. An additional local peak in $R_{xx}$ can be identified at $v_h = 3/4$. Figure 3c-f show the symmetrized $R_{xx}$ and anti-symmetrized $R_{xy}$ as a function of $v_h$ at representative $E$ and $B$-fields.

At $v_h = 1$, a strong dip in $R_{xx}$ together with quantized $R_{xy}$ at $\frac{h}{e^2}$ can be identified, corresponding to the QAH state at $v_h = 1$, consistent with previous studies[3,4]. Intriguingly, at $v_h = 1/2$, a local dip in $R_{xx}$ together with quantized $R_{xy}$ at $\frac{h}{e^2}$ (Fig. 3a, 3b, 3d, 3f and 3g) can be identified within a certain $E$ range, signifying $v_h = 1/2$ is a gapped state with an integer Chern number. This is in sharp contrast with the twist bilayer MoTe$_2$, where a gapless liquid phase, likely CFL, is observed at $v_h = 1/2$[3,4,23]. At $v_h = 2/3$, we also observe a local dip in $R_{xx}$ together with a quantized $R_{xy}$ plateau at $\frac{h}{e^2}$ that emerges within a certain $E$ range (Fig. 3a-c), signifying $v_h = 2/3$ is also a gapped state with an integer Chern number. This is also different from the twist bilayer MoTe$_2$, where the FQAH phase is observed at $v_h = 2/3$[3,4]. While quantized $R_{xy}$ is observed at nearly zero magnetic field for $v_h = 1$ and $1/2$, a finite field of

~0.5 T is required to achieve quantization at $v_h = 2/3$ (Fig. 4a and 3c). In addition to $v_h = 1, 2/3$ and $1/2$, another quantized Hall plateau and corresponding $R_{xx}$ dip are observed at $v_h \sim 0.53$ within a certain $E$-field range (Fig. 3a, 3b, 3d, 3e and 3g). The $v_h \sim 0.53$ state persists down to zero magnetic field and exhibits magnetic hysteresis. Although very close to $v_h = 1/2$, they are separated by a dissipative metallic state with large $R_{xx}$ and suppressed Hall (Fig. 3a, 3b and 3d), showing they are different states.

The Chern numbers of these states can be further examined by measuring $R_{xx}$ and $R_{xy}$ as a function of $v_h$ and $B$. As shown in Figures 4a and 4b, $v_h = 1$ is featured by a broad local $R_{xx}$ dips that show a linear shift in $v_h$ with increasing $B$. The dispersion agrees well with $C = 1$ according to the Streda formula $n_M \frac{dv_h}{dB} = C \frac{e}{h}$, as illustrated by the dashed lines. For $v_h = 2/3$, the dispersion is better resolved in the $R_{xy}$ map, which agrees well with $C = 1$, as illustrated by the dashed lines (Fig. 4a). For the $v_h = 1/2$ state, the dispersion is better resolved in the $R_{xx}$ map, which agrees reasonably well with $C = 1$, as illustrated by the dashed lines (Fig. 4b). Surprisingly, the $v_h \sim 0.53$ state shows an abnormally strong dispersion under small out-of-plane $B$-field (<0.5 T), far exceeding $C = 1$ expectation based on its Hall signal (Extended Data Fig. 9), whose origin remains open. The slope decreases dramatically at $B$-field larger than 0.5 T, where the $v_h \sim 0.53$ state is gradually suppressed. While excellent quantization with residual $R_{xx}$ of $\sim 100\ \Omega$ is observed at low field ~0.1 T (Fig. 3g), $R_{xy} \approx 0.84 \frac{h}{e^2}$ and $R_{xx} \approx 15\ k\Omega$ is observed at $B = 2\ T$ (Fig. 3f). For $B > 5T$, the $v_h \sim 0.53$ state vanishes completely (Fig. 4b and Extended Data Fig. 6). The anomalously strong dispersion and $B$-field dependence suggests an unconventional underlying mechanism, which requires further theoretical and experimental investigations.

**Discussion**

In general, the formation of a gapped topological state at a commensurate fractional filling can occur in two scenarios: FQAH and TEC. These two scenarios generically compete with each other: while the first dominates in A-A $t$MoTe$_2$, introducing the B layer switches the system into the second. Based on the extracted electrostatic phase diagram, a finite charge transfer gap exists between the first moiré band and the B-layer band. Thus, holes only occupy the first moiré band, and the B layer remains neutral when the TEC states emerge, illustrating that the influence of the B layer cannot be understood by the charge transfer.

Recent theory predicts that $t$MoTe$_2$ can host both FQAH and TEC phases, where the exact ground state is sensitive to the quantum geometric condition of the band[17,18,26]. We propose that the TEC states observed at $v_h = 1/2$ and $2/3$ arise from the modified quantum geometry of the first moiré band by the B layer. We consider that the presence of the B-layer introduces an extra intralayer potential $V_B$, originating from the moiré-periodic lattice relaxation in the B layer. This intralayer potential redistributes Berry curvature and increases the variance of quantum metric elements across the Brillouin zone, driving the band geometry further away from the ideal limit. This picture is supported by our calculation of the trace condition of the first moiré band (Extended Data Fig. 8f), which shows that the trace condition is suppressed by $V_B$. Such quantum geometric non-ideality disfavors delocalized fractional quasiparticles, promoting real-space symmetry breaking and electronic crystallization while preserving overall band topology ($C = 1$).

Band-projected exact diagonalization (Fig. 3h) further reveals that, in the AA-bilayer limit, the ground-state manifold is quasi-sixfold degenerate for $N_s = 28$, characteristic of a CFL. However, inclusion of the B-layer potential results in a quasi-fourfold degenerate ground state with a many-body Chern number $C = 1$, consistent with a TEC and in agreement with experimental observations at $\nu_h = 1/2$ in A-AB $t$MoTe$_2$. Thus, a remote layer serves as a practical quantum geometry tuning knob—complementary to twist-angle, pressure, or strain modulation.

Notably, although the generalized Wigner crystals have been extensively studied in TMD moiré systems such as WSe$_2$/WS$_2$ heterobilayers[37,38], and although the TEC phases with QAH[9,29–31] have been observed in graphene moiré systems, it is the first time TEC has been observed in a TMD moiré system. The $\nu_h = 1/2$ TEC may further break rotational symmetry (possible stripe-like charge order), testable by future anisotropic transport, optical birefringence[39], scanning probe microscopy[40,41], or momentum-space diffraction[42].

In contrast, the observed integer-QAH state at $\nu_h \approx 0.53$ is distinct from the TEC states at $\nu_h = 1/2$ and $2/3$. Although both characterized by integer QAH at fractional filling factors, TEC states occur exclusively at commensurate moiré fillings, while $\nu_h \sim 0.53$ is incommensurate. This state also differs from the recently observed extended quantum anomalous Hall (EQAH) effect in rhombohedral graphene moiré systems[11], where integer quantized anomalous Hall conductance persists over a broad moiré filling factor range. A possible explanation for the $\nu_h \sim 0.53$ state is the lattice analogue and zero-field version of the reentrant integer quantum Hall (RIQH) effect observed in Landau levels[43–49], i.e., the reentrant integer quantum anomalous Hall (RIQAH) effect, which has also been observed very recently in high-quality twisted bilayer MoTe$_2$ at $\nu_h \sim 0.63$[50]. In a 2DEG system under strong magnetic fields, the RIQH state is one type of electron solid that competes with the fractional quantum Hall liquid. The RIQH effect occurs at fractional fillings (not necessarily commensurate) of Landau levels and can generally be understood as the superposition of a weakly pinned Wigner crystal (WC) and a filled Landau level[51]. Notably, the RIQH state at $\nu_h \approx 0.53$ has never been observed in the lowest Landau level. The other possible candidate is the QAHC phase proposed recently[52–55]. The $\nu_h \sim 0.53$ is an incommensurate filling factor. Therefore, the role of moiré potential in the $\nu_h \sim 0.53$ state may differ from that in TEC states. Further experiments, such as imaging the electronic lattice in real space[40,41], or momentum space[42], or noise spectrum measurement[56–58] are needed to clarify its nature, which is beyond the scope of this study.

Our study provides experimental evidence of the competition and switching between

CFL/FQAH and TEC at fractional fillings of a flat Chern band driven by quantum geometry modification, demonstrating controllable competition between fractionalization and symmetry breaking. Our experiment also paves the way for exploring correlation-driven topological phenomena by tuning quantum geometry.

**Note added:** During the preparation of this manuscript, we became aware of another theoretical work addressing a similar topic[36].

## Methods

### Device fabrication

Devices were fabricated using the standard tear-and-stack and dry transfer method. In all devices, to facilitate good contact formation for hole doping in $t$MoTe$_2$, we use few-layer TaSe$_2$ as contact electrodes, taking advantage of its high work function. Thin flakes of hexagonal boron nitride ($h$BN), graphite ($3-5$ nm), monolayer MoTe$_2$ and WSe$_2$ (HQ Graphene), and TaSe$_2$ ($2-5$ nm) were first exfoliated from bulk crystals onto Si/SiO$_2$ (285 nm) substrates. Flakes of appropriate thickness were identified according to their reflectance contrast under an optical microscope.

We selected exfoliated MoTe$_2$ flakes with both monolayer and bilayer regions to ensure a precise twisted angle between them. The flake was mechanically cut into two parts using AFM tips in the monolayer region, then stacked together to get both monolayer-bilayer and monolayer-monolayer regions. Two stacking orders, i.e., A-AB and A-BA, exist naturally, while the A-AB structure was selectively studied by picking up the bilayer part first (Extended Data Fig. 1a). We handled MoTe$_2$ flakes inside a nitrogen-filled glovebox with oxygen and water levels below $0.01$ ppm to minimize the degradation of MoTe$_2$. The thickness of $h$BN flakes was also measured using an atomic-force microscope (AFM). The complete dual-gated structure was assembled using a polycarbonate stamp within the N$_2$ glovebox environment. A high-precision rotation stage (Thorlabs PRM1Z8) controls the targeted twist angle with a typical accuracy of $\sim 0.1°$. To achieve ohmic contacts irrespective of the applied electric fields, we fabricated the dual-gated devices with SiO$_2$/Si global contact gates to heavily hole-dope the contact regions, where $t$MoTe$_2$ overlaps with few-layer TaSe$_2$ flakes (Extended Data Fig. 1b). To define the Hall bar geometry of the $t$MoTe$_2$ channel and eliminate the unavoidable monolayer MoTe$_2$ region, we carried out extra EBL and RIE processes, following a procedure similar to the one reported earlier[4]. The dual-gate geometry allows independent control of $v_h$ and $E$ by the top gate voltage $V_t$ and bottom gate voltage $V_b$.

## Optical measurements

The optical measurements were performed in reflection geometry in a confocal optical microscope system based on a closed-cycle helium cryostat (Attodry2100, base temperature 1.6 K) equipped with a 9 T superconducting magnet. A superluminescent light-emitting diode with a peak wavelength of 1070 nm and a full-width at half-maximum bandwidth of 90 nm was used as the light source. The output of the diode was coupled to a single-mode fiber and focused onto the device under normal incidence by a low-temperature microscope objective (Attocube, 0.8 numerical aperture). The spot size of the excitation light, defined as the FWHM of the diffraction-limited beam spot, can be calculated as $\text{FWHM} = 0.51 \frac{\lambda}{\text{NA}} \sim 0.7$ μm. A combination of a linear polarizer and an achromatic quarter-wave plate was used to generate the left- and right-circularly polarized light. To avoid perturbing the system, the incident intensity on the sample was kept below 5 nW μm$^{-2}$. The reflected light of a given helicity was spectrally resolved by a spectrometer coupled to a liquid nitrogen-cooled InGaAs one-dimensional array sensor (Princeton Instruments PyLoN-IR 1.7). The reflectance contrast spectrum was obtained by comparing the reflected light spectrum from the sample with the reference spectrum measured on a heavily doped condition (which is featureless in the spectrum of interest).

The MCD spectrum is defined as $\text{MCD}(E) = \frac{R^+(E) - R^-(E)}{R^+(E) + R^-(E)}$, where $R^+(E)$ and $R^-(E)$ denote the reflection intensity spectrum of the left- and right-circularly polarized light. To analyze the MCD as a function of tuning parameters such as $\nu$, $E$, $B$, and $T$, we integrate the MCD modulus over a spectral range that covers most of the spectrum range of the MCD signal. The integrated MCD (referred to simply as the MCD below) reflects the difference in occupancy between the K and K′ valleys in MoTe$_2$. Because of spin-valley locking, the signal is proportional to the out-of-plane magnetization[35,35,59]. The MCD maps are anti-symmetrized by positive and negative $B$, i.e., $\text{MCD} = \frac{\text{MCD}(+B) - \text{MCD}(-B)}{2}$.

In the PL measurements, a 632.8-nm helium–neon laser was used with excitation power around 50 nW. A 650 nm long-pass filter was inserted before the entrance to the spectrometer to reject the reflected laser excitation.

## Transport measurements

Electrical transport measurements above 1.5 K were performed in a closed-cycle $^4$He cryostat (Oxford TeslatronPT) equipped with a 12 T superconducting magnet. Measurements below 1.5 K were performed in a top-loading dilution refrigerator (Oxford TLM, nominal base temperature about 15 mK) equipped with an 18 T superconducting magnet. The sample was immersed in the $^3$He–$^4$He mixtures during the measurements.. Each line of the dilution refrigerator has a silver epoxy filter and an RC filter (consisting of a 470-Ω resistor and a 100-pF capacitor) at low temperatures. Electrical transport measurements were conducted using standard low-frequency lock-in techniques. The bias current was limited within 2 nA to avoid sample heating and disturbance of fragile quantum states. Voltage preamplifiers with 100 MΩ impedance were used to improve the signal-to-noise ratio. Finite longitudinal-transverse coupling occurs in our devices that mixes the longitudinal resistance $R_{xx}$ and Hall resistance $R_{xy}$. To correct this effect, we used the standard procedure to symmetrize $\left[\frac{R_{xx}(B)+R_{xx}(-B)}{2}\right]$ and anti-symmetrize $\left[\frac{R_{xy}(B)-R_{xy}(-B)}{2}\right]$ the measured $R_{xx}$ and $R_{xy}$ under positive and negative magnetic fields to obtain accurate values of $R_{xx}$ and $R_{xy}$, respectively. The measurements were carried out without any optical excitation.

**Determination of doping density and electric field**

Here, we define the applied electric field in vacuum as $E = D/\varepsilon_0$, where $D$ is the displacement field and $\varepsilon_0$ is the vacuum permittivity. The carrier density $n$ and $E$ on the sample are converted from the top (bottom) gate voltage $V_t$ ($V_b$) using a parallel plate capacitor model:

$$n = \frac{V_t C_t + V_b C_b}{e} = \frac{1}{e}\left(\frac{\varepsilon_0 \varepsilon_t}{d_t}V_t + \frac{\varepsilon_0 \varepsilon_b}{d_b}V_b\right)$$

$$D = \frac{V_t C_t - V_b C_b}{2} - D_{\text{offset}} = \frac{1}{2}\left(\frac{\varepsilon_0 \varepsilon_t}{d_t}V_t - \frac{\varepsilon_0 \varepsilon_b}{d_b}V_b\right) - D_{\text{offset}}$$

where $e$ is the electron charge, and $C_t$ and $C_b$ are the top and bottom gate capacitances obtained from the top and bottom $h$BN thickness $d_t$ and $d_b$, which can be measured by atomic force microscopy (Asylum Research Cypher S). The offset electric field likely arises from the structure asymmetry and can be determined from the symmetric axis of the dual-gate MCD map. In principle, $n$ can be obtained from the capacitor model and can then be used to calculate the twist angle from the assigned filling factors. In our case, thanks to the high mobility of our device, $n$ (and thus, the twist angle) can be independently calibrated from the $R_{xx}$ map and Landau fan diagrams with better accuracy. The Extended Data Fig. 2 presents the $B$ and filling dependences of $R_{xx}$. The Landau fan diagram, denoted by dashed lines, highlights a series of minima in $R_{xx}$ corresponding to the different Landau levels. Upon complete filling of $\nu_{LL}$th Landau levels with hole carriers, the hole carrier density can be expressed by $n_{\text{SdH}} = \nu_{LL}|B|/\phi$, where $\phi = h/e$ ($h$ is Planck's constant). By determining the number of filling factors corresponding to $n_{\text{SdH}}$, moiré density $n_M$ can be obtained. In the experiment, we calculated $n_{\text{SdH}}$ at several fixed $B$, and plotted them as a function of $\nu_h$. From the slope of the linear fit, we obtained a moiré density $n_M = (3.90 \pm 0.15) \times 10^{12}$ cm$^{-2}$. Finally, the twist angle $\theta$ was calculated to be $3.71° \pm 0.08°$ using $\theta = a\sqrt{\frac{\sqrt{3}}{2}n_M}$, where $a$ denotes the MoTe2 lattice constant, approximately 3.5 Å.

**Electrostatic model for interlayer excitons in TMD bilayers**

To obtain the effective out-of-plane electric dipole $D_e$ of interlayer excitons, we calculate the local electric field $E_M$ within the MoTe2 layer for a given applied field $E$, using the parallel-plate capacitor model[60,61]:

$$E_M \approx \frac{\epsilon_{hBN}}{\epsilon_M} E$$

Here $\epsilon_{hBN}$ and $\epsilon_M$ are the local effective out-of-plane dielectric constants for the hBN and MoTe2 layers, respectively. For a fixed doping density $n$, the Stark shift $E_{iX}$ for interlayer excitons becomes

$$E_{iX} = et \times E_M \approx et \frac{\epsilon_{hBN}}{\epsilon_M} E = D_e E$$

where $D_e = \frac{\epsilon_{hBN}}{\epsilon_M} et$ is the effective out-of-plane electric dipole and is dependent on the local dielectric environment of the moiré superlattice. The effective electric dipole $D_e \approx 0.17\ e \cdot$nm can be estimated, using the typical parameters $\epsilon_{hBN} \approx 3.7$, $\epsilon_M \approx 9$[62], and $t \approx 0.7$ nm.

**The continuum model of A-AB tMoTe2**

To enable our many-body simulations, we build a continuum model of the system. Owing to the large momentum separation between the two valleys and the strong Ising spin–orbit coupling near the band edge, we focus on the K valley and obtain the continuum model below:

$$H(k) = \begin{bmatrix} -\frac{\hbar^2 k^2}{2m^*} + V_{A+}(r) & T(r) & 0 \\ T^*(r) & -\frac{\hbar^2 k^2}{2m^*} + V_{A-}(r) + V_B(r) & 0 \\ 0 & 0 & -\frac{\hbar^2 k^2}{2m^*} + V_B(r) \end{bmatrix}$$

Where:

$$V_B = \sum_{i=1,3,5} 2V_B \cos(G_i \cdot r + \phi)$$

$$V_\pm = \sum_{i=1,3,5} 2V_1 \cos(G_i \cdot r \pm \phi_1)$$

$$T = \sum_{i=1,2,3} w_1 e^{-iq_i \cdot r}$$

$G_i$ is a moiré reciprocal lattice vector, $q_i$ denotes the momentum differences between nearest plane-wave bases across different layers. $V_{A+}(V_{A-}), V_B$ represent the layer dependent intralayer potential in bottom (top) A layer and the nearby B layer; $T(r)$ indicates the interlayer moire potential between the AA layer. Due to the nearby pocket of aligned AB layer carrying opposite spin due to the Ising SOC, we ignore the interlayer coupling between the aligned AB layers. In all of the calculations, we use the parameter: $\theta = 3.7°$, $V_1 = 11.2 \text{ meV}$, $\phi_1 = -91°$, $w_1 = -13.3 \text{ meV}$, $\phi = 120°$. Besides, we use the dual-gate Coulomb potential, and the screened length is chosen to be 10 nm.

**Acknowledgments**
We thank Quansheng Wu, Jiabin Yu and Ziyue Qi for fruitful discussions. Research was primarily supported by the National Key R&D Program of China (Nos. 2021YFA1401400, 2021YFA1400100, 2022YFA1405400, 2022YFA1402702, 2022YFA1402404, 2019YFA0308600, 2022YFA1402401, 2020YFA0309000), the National Natural Science Foundation of China (Nos. 12174250, 12141404, 12350403, 12174249, 92265102, 12374045), the Innovation Program for Quantum Science and Technology (Nos. 2021ZD0302600 and 2021ZD0302500), the Natural Science Foundation of Shanghai (No. 22ZR1430900). S.J., T.L., X.L. and S.W. acknowledge the Shanghai Jiao Tong University 2030 Initiative Program B (WH510207202). Z.C. acknowledges the support of NSAF (No. U1530402). T.L. and S.J. acknowledge the Yangyang Development Fund. K.W. and T.T. acknowledge support from the JSPS KAKENHI (Grant Numbers 21H05233 and 23H02052) , the CREST (JPMJCR24A5), JST and World Premier International Research Center Initiative (WPI), MEXT, Japan. Device fabrication is supported by the Micro-nano Fabrication Platform of the School of Physics and Astronomy at Shanghai Jiao Tong University.

**Author contributions**
S.J. and T.L. designed the scientific objectives and oversaw the project. F.L., F. X., Z.S. and X.C. fabricated the devices, with the help of X.T.. F. X., F. L. performed the transport measurements, with the help of J.L.. F. L and J.X. performed the optical measurements. F.L., F.X., J.X., T.L. and S.J. analyzed the data. C.X., N.M. and Y.Z. performed theoretical studies. R.Z. grows the bulk $TaSe_2$ crystals. K.W. and T.T. grew the bulk hBN crystals. F.L., Y.Z., T.L. and S.J. wrote the manuscript. All authors discussed the results and commented on the manuscript.

# Figures

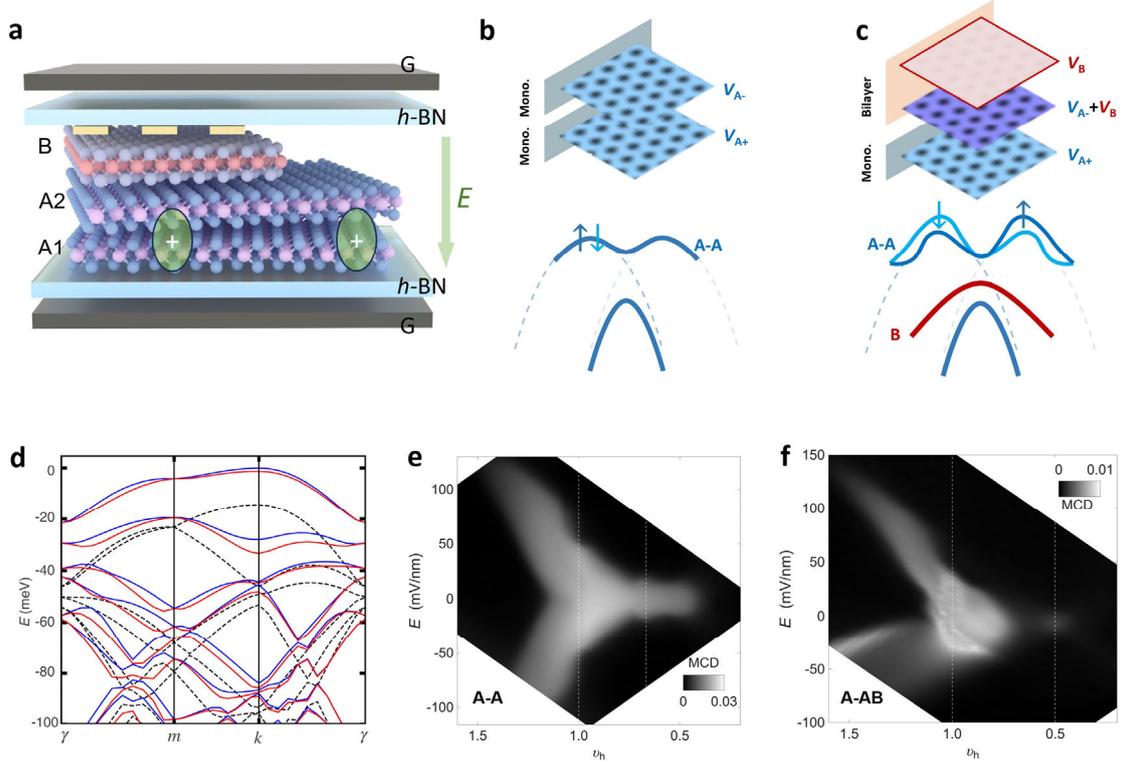

**Fig. 1 | Magnetic characterization of A-A and A-AB *t*MoTe$_2$.**
**a,** Schematic of the *t*MoTe$_2$ device consisting of both A-A and A-AB *t*MoTe$_2$ with the same twist angle. Rotationally misaligned monolayer (A1, bottom) and AB bilayer MoTe$_2$ (B+A2, top) are sandwiched between two graphite/*h*BN gates, enabling independent control of both $E$ and $n$ (Si contact gate is not shown). The green arrow indicates the positive direction of $E$-field. **b,c,** Upper panels: Schematic representations of the real-space MoTe$_2$ layers of A-A and A-AB *t*MoTe$_2$, with the rotated interface between layers with different color projections, i.e., gray for monolayer and yellow for bilayer. $V_{A+}$ ($V_{A-}$), $V_B$ represent the layer dependent intralayer potential in bottom (top) A layer and the nearby B layer. Lower panels: Schematic representation of the corresponding band structures in momentum space, with blue bands from the A-A bilayer and the red from the B layer. The energy dispersion of the first moiré band is modified by the B layer, and the valley/spin degeneracy is broken. **d,** First-principal band structure of A-AB *t*MoTe$_2$. Here, we neglect the interlayer coupling between the A-A bilayer and the B layer. The red (blue) curves show the spin-up (spin-down) bands calculated for the A-A bilayer extracted from the relaxed A-AB *t*MoTe$_2$, while the black dashed curves show the bands of the isolated B layer taken from the same structure. **e,f,** MCD maps as a function of $v_h$ and $E$ of A-A and A-AB regions, respectively, under a small $B$ (10 mT). $v_h = 1$, $2/3$, and $1/2$ are indicated by the vertical dashed lines. Measurements were performed at 1.6 K.

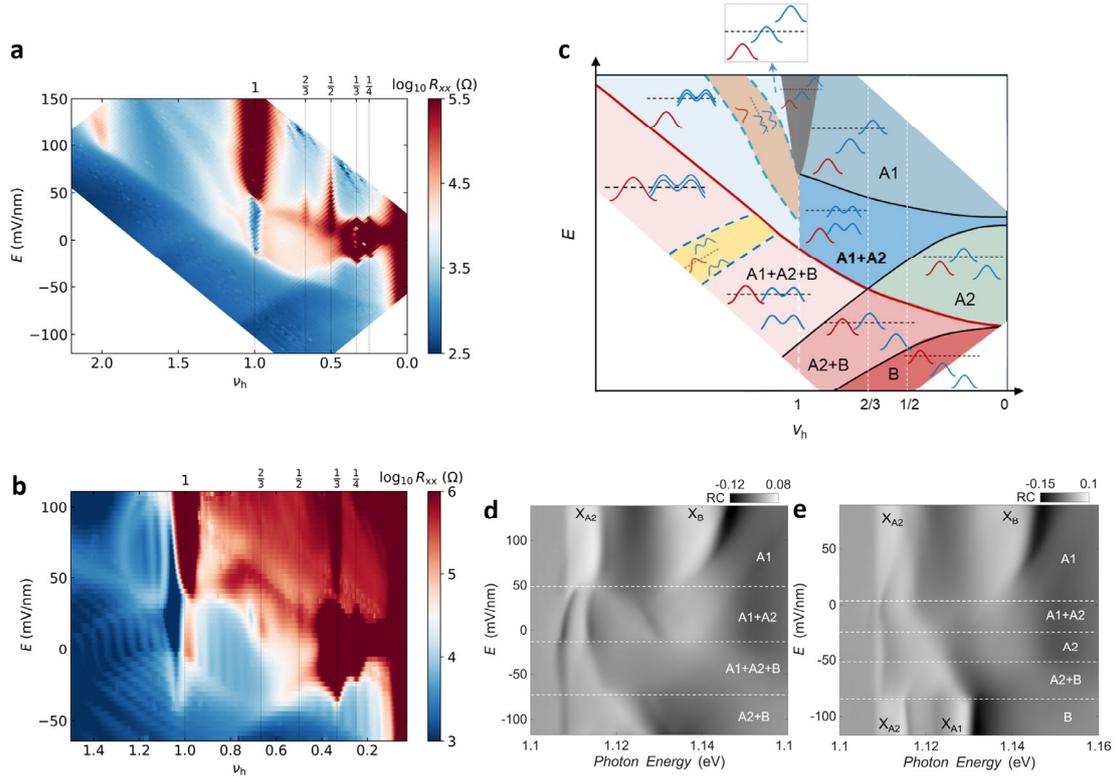

**Fig. 2 | Electrostatic phase diagram of A-AB *t*MoTe₂.**

**a,b,** $R_{xx}$ as a function of $E$ and $v_h$ at 1.5 K, symmetrized at $B = \pm 0.1$ T and $\pm 10$ T, respectively. $v_h = 1, 2/3, 1/2, 1/3$ and $1/4$ are indicated by the vertical dashed lines. **c,** $(v_h, E)$ of A-AB *t*MoTe₂, where regions are defined by different layer polarization configurations. The regions are denoted by the (combination of) layer numbers (definition in Fig. 1a), corresponding to which layers are occupied. The inset in each region schematically illustrates the band structure, with blue bands from the A-A bilayer and red from the B layer. The double-peak bands represent the hybridized A-A band, while the single-peak bands represent the non-hybridized band. The Fermi level is denoted by the black dashed line. The layer is charge-neutral if its band lies below the Fermi level. Note that the Mott gap will vanish at filling factors significantly larger than 1. **d,e,** Optical reflectance contrast (RC) spectrum as a function of $E$ at $v_h = 1$ and $1/2$, respectively. The horizontal dashed lines separate different layer-polarized regions. The strong resonance features at large $E$-field correspond to neutral excitons (X) of different layers, denoted as $X_{A1}$, $X_{A2}$ and $X_B$. For $v_h = 1$, the threshold $E$-field to reach region B is too large, which is beyond the capability of the gates.

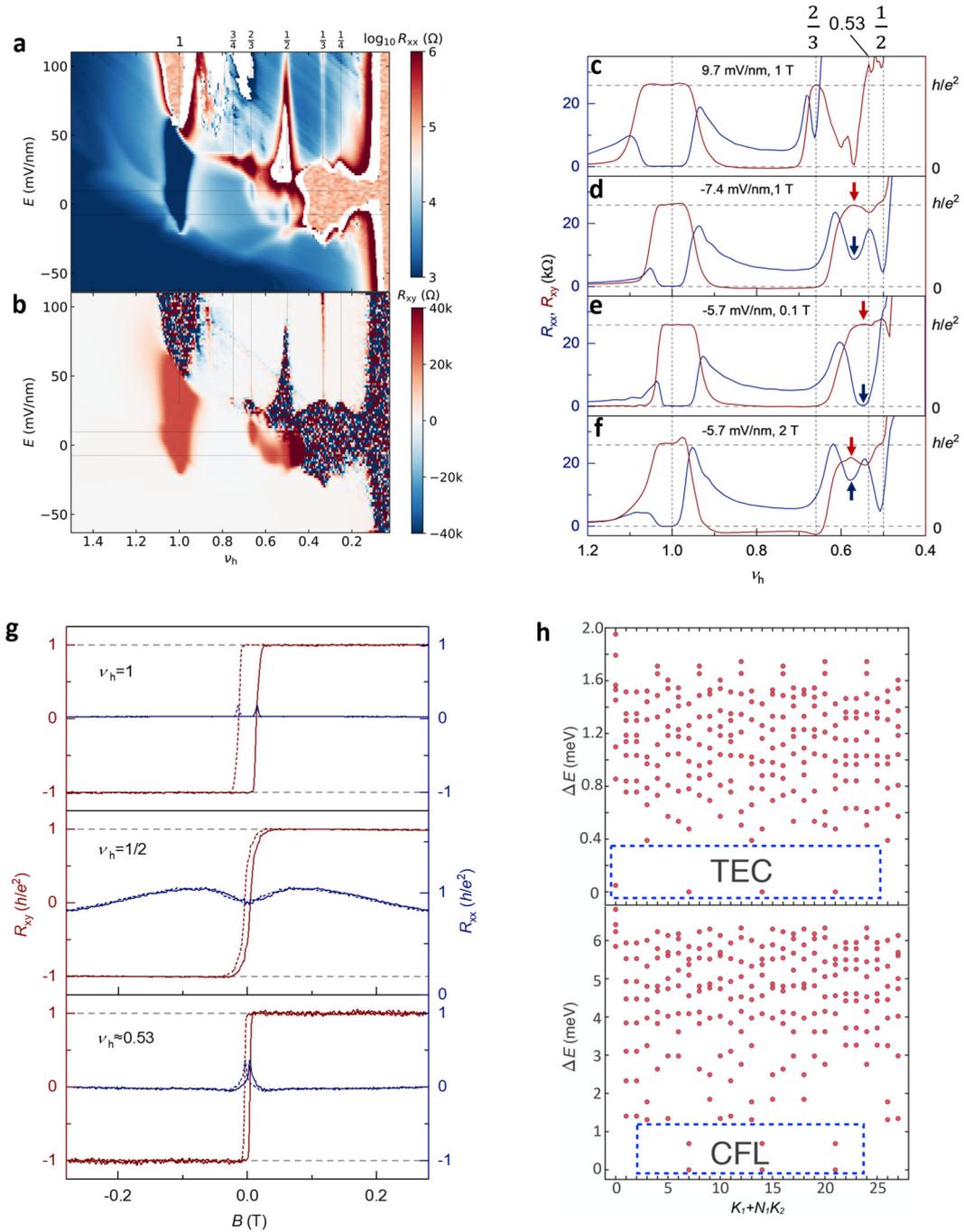

**Fig. 3 | TEC states at commensurate and incommensurate fractional filling factors.**
**a,b,** $R_{xx}$ and $R_{xy}$ as a function of $E$ and $\nu_h$ at 15 mK, symmetrized (anti-symmetrized) at $B = \pm 1$ T, respectively. $\nu_h = 1, 3/4, 2/3, 1/2, 1/3$ and $1/4$ are indicated by the vertical dashed lines. Horizontal dashed lines denote $E = 9.7$ and $-7.4$ mV nm$^{-1}$. **c-f,** $R_{xy}$ and $R_{xx}$ as a function of $\nu_h$ at representative $E$ and $B$-fields. The quantized values are indicated by the horizontal dashed lines. $\nu_h = 1, 2/3, 0.53$, and $1/2$ are indicated by the vertical dashed lines. The arrows indicated the $\nu_h \sim 0.53$ state at finite $B$-field. Note its significant filling factor shift under $B$-field due to its abnormally large dispersion (Extended Data Fig. 9). While quantized $R_{xy}$ plateau is observed at $B = 0.1$ T, the $\nu_h \sim 0.53$ state is suppressed by $B$, and $R_{xy}$ deviates significantly from the quantized value at $B = 2$ T. **g,** $B$ dependence of $R_{xy}$ and $R_{xx}$ at $\nu_h = 1, 1/2$ and $0.53$ at $E = -7.4$ mV nm$^{-1}$. The quantized values are indicated by the dashed lines. **h,** Top: band-projected exact diagonalization spectrum of the A-AB $t$MoTe2 (continuum model, system size $N_s = 28$) at $v_h = 1/2$ ($\theta = 3.70°$, $V_B = 1$ meV) yields a four-fold quasi-degenerate ground-state manifold, where $\Delta E$ is the energy difference with respect to the energy of ground states. $K_1 + N_1 K_2$ is used to index the kpoints in the kmesh (details can be found in Methods and Extended Data Fig. 8e). Each state carries a many-body Chern number $C = 1$, consistent with TEC. Bottom: With the same twist angle ($\theta = 3.70°$, $V_B = 0$ meV), the calculation produces a six-fold quasi-degenerate ground-state manifold, characteristic of the CFL.

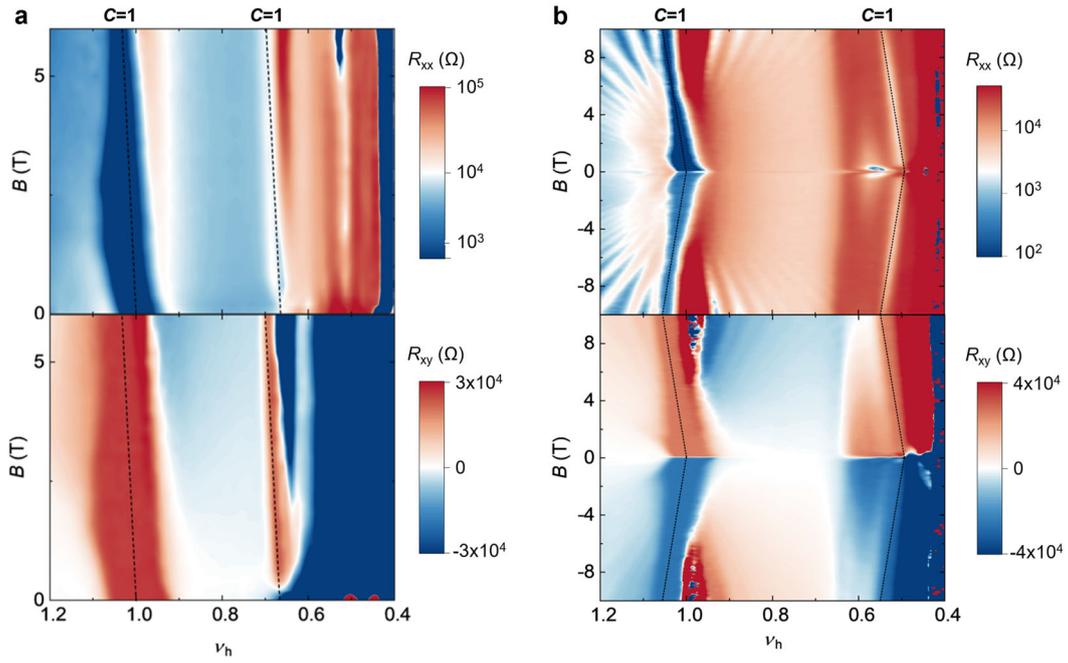

**Fig. 4 | Magnetic field dependence.**

**a,** $R_{xx}$ and $R_{xy}$ versus out-of-plane $B$ and $\nu_h$ at $E = 6.3$ mV nm$^{-1}$. Dashed lines represent the expected dispersions based on the Streda formula for $\nu_h = 1$ and $2/3$ with $|C| = 1$. **b,** $R_{xx}$ and $R_{xy}$ versus out-of-plane $B$ and $\nu_h$ at $E = -5.7$ mV nm$^{-1}$. Dashed lines represent the expected dispersions based on the Streda formula for $\nu_h = 1$ and $1/2$ with $|C| = 1$. All the measurements are performed at 15 mK.

# Extended Data Figures

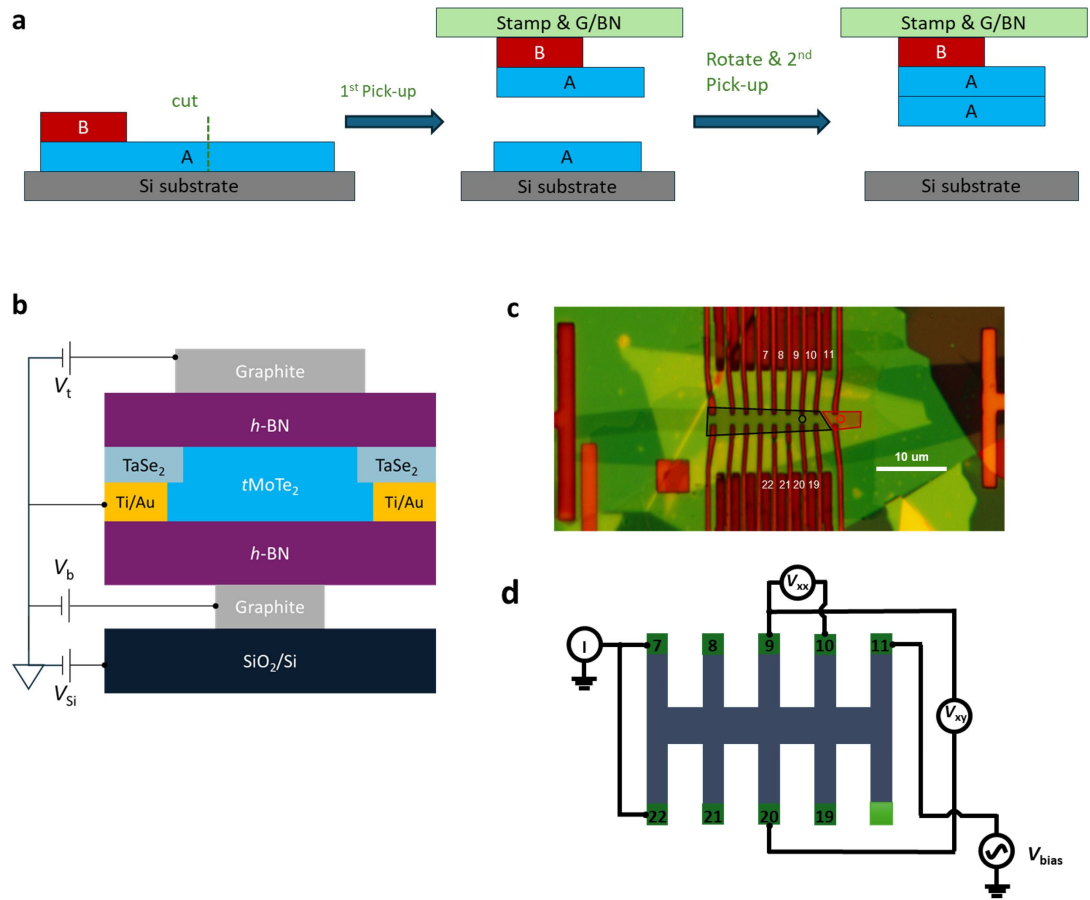

**Extended Data Figure 1 | Device geometry.**
**a,** Schematics of the key fabrication process of our device consisting of both A-A and A-AB $t$MoTe$_2$ regions. Exfoliated MoTe$_2$ flakes with both monolayer and bilayer parts are selected, and the 'cut and stack' method is used to ensure the A-A and A-AB regions have the same twist angle. **b,** Schematic of a dual-gated device with contact gate design for optical and transport measurements. $V_\mathrm{t}$, $V_\mathrm{b}$ and $V_\mathrm{si}$ are the bias voltages applied to the top/bottom graphite/$h$BN gate, and the SiO$_2$/Si global contact gate, respectively. **c,** Optical micrograph of the twisted monolayer-bilayer MoTe$_2$ device used in this study. The scale bar is 10 μm. The Hall bar geometry is defined by standard EBL and RIE processes. The red and black lines outline the boundaries of the twisted A-A region and the twisted A-AB region, respectively. Optical measurement data are taken at the red and black circle positions. **d,** Schematic figure of the measurement configuration. For most results presented in the main text, electrodes 7 and 22 are grounded, and electrode 11 is used as a source electrode. The longitudinal voltage drop is measured between 9 and 10, and the transverse voltage drop is measured between 9 and 20.

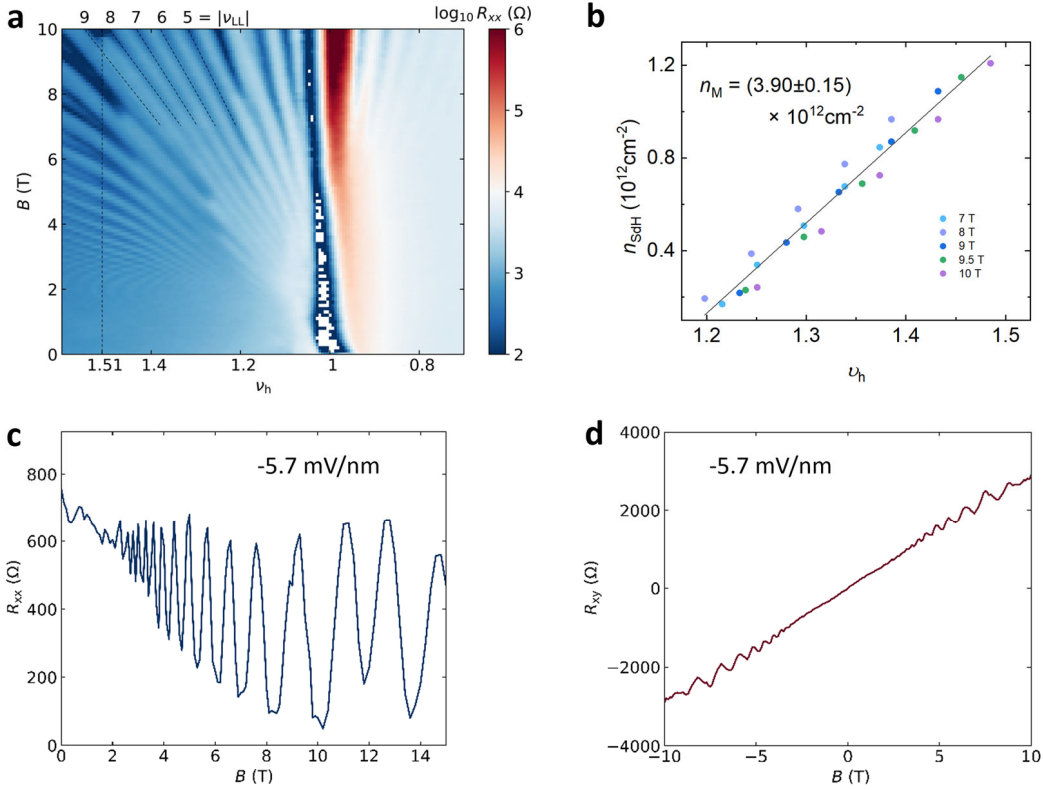

**Extended Data Figure 2 | Twist angle calibration and mobility estimation.**
**a,** $R_{xx}$ versus out-of-plane $B$ and $\nu_h$ at $E = -5.7$ mV nm$^{-1}$ at $T$=15 mK for A-AB $t$MoTe$_2$ device. **b,** At several fixed B, we calculate the carrier density based on Shubnikov–de Haas (SdH) oscillations using $n_{SdH} = \nu_{LL} e|B|/h$, and plot them as a function of $\nu_h$. From the slope of the linear fit, we obtain a moiré density $n_M = (3.90 \pm 0.15) \times 10^{12}$ cm$^{-2}$. The twist angle is determined to be 3.71°±0.08°. **c,** Representative $R_{xx} - B$ curve, a line cut of **a**, as denoted by the vertical dashed line, at $\nu_h = 1.51$. Well-resolved SdH oscillations emerge at $B$ above ~2 T. From the criterion of $\mu_q B = 1$ for the onset of SdH oscillations, a quantum mobility of $\mu_q \sim 5000$ cm$^2$/Vs is obtained. **d,** Corresponding $R_{xy} - B$ curve at $\nu_h = 1.51$. The zero-field longitudinal resistance, $R_{xx} = 754$ Ω, is obtained from **c**. The coefficient $\frac{1}{n_{2D}e} = 322$ Ω/T is extracted through linear fitting of $R_{xy} - B$ within ±1 T range. The Hall mobility can be calculated using the relationship $\mu_H = \frac{L}{W} \frac{1}{R_{xx} n_{2D} e}$, where $L$=2.2 μm and $W$=1.2 μm are the length and width of the relevant probes, respectively. $\mu_H \sim 7800$ cm$^2$/Vs is extracted, in reasonable agreement with $\mu_q$. The obtained $\mu_q$ and $\mu_H$ are the highest value in the $t$MoTe$_2$ moiré devices so far, demonstrating the high quality of our device.

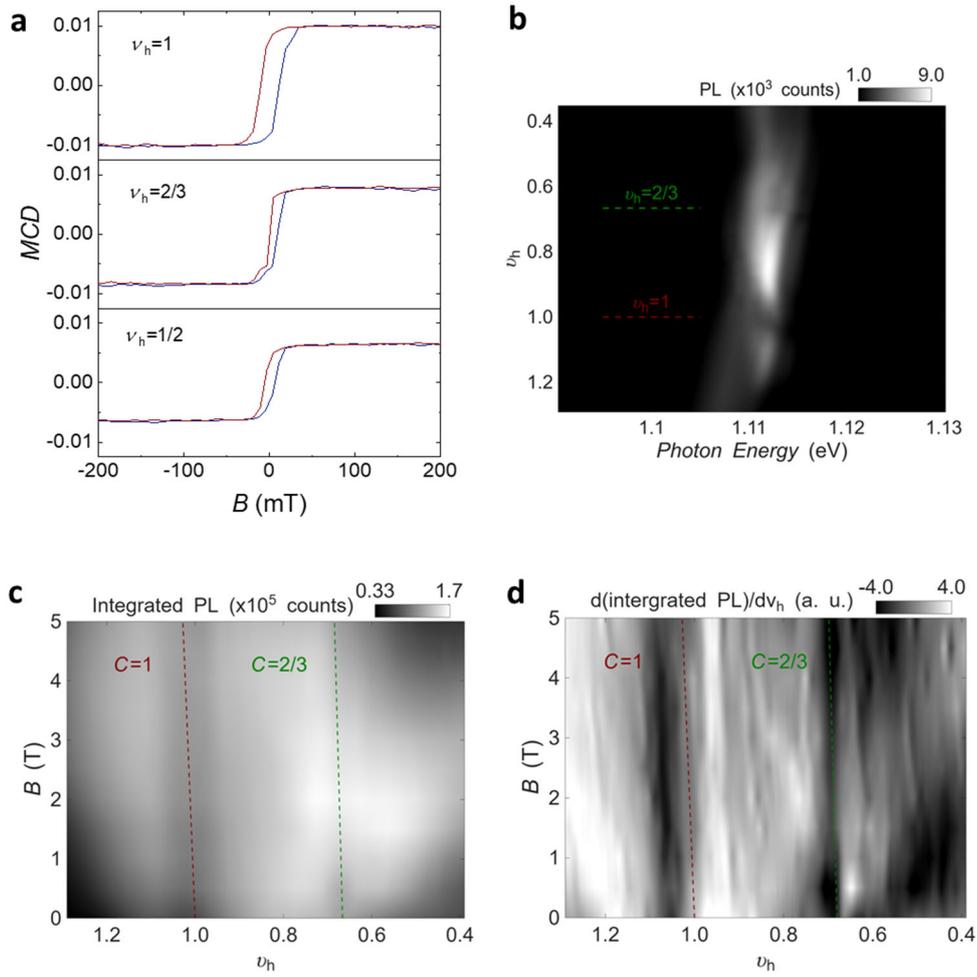

**Extended Data Figure 3 | Optical characterizations of the A-A region.**

**a,** MCD as a function of $B$ at representative filling factors $v_h = 1,\ 2/3$ and $1/2$. **b,** PL intensity plot as a function of hole doping and photon energy. Filling factors $v_h$ corresponding to the formation of correlated insulating states are indicated. **c,d,** Spectrally integrated PL intensity and its derivative versus $B$ and $v_h$, respectively. Dashed lines represent the expected dispersions based on the Streda formula for $v_h = 1$ and $2/3$ with $|C| = 1$ and $2/3$, respectively.

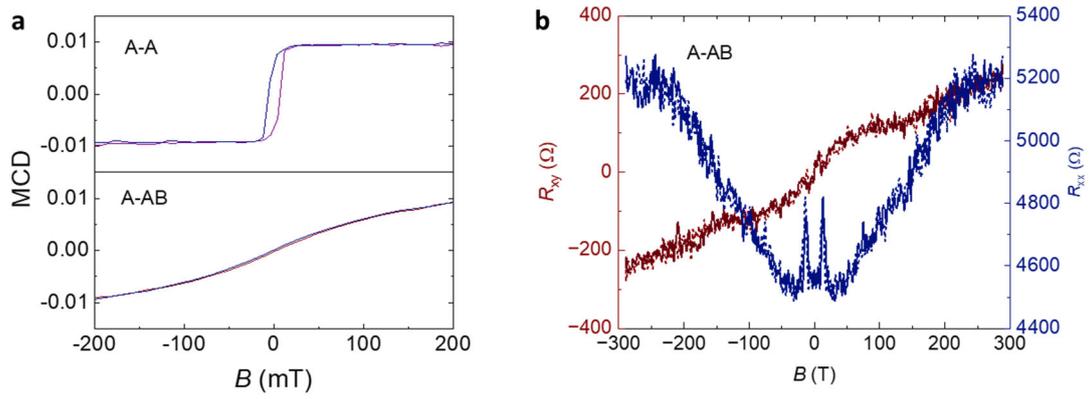

**Extended Data Figure 4 | Characterization of magnetic property at $v_\mathrm{h} = 0.7$.**

**a,** MCD as a function of $B$ at $v_\mathrm{h} = 0.7$ on A-A and A-AB regions, measured at 1.6 K. **b,** $B$ dependence of $R_{xy}$ and $R_{xx}$ of A-AB at $v_\mathrm{h} = 0.7$ and $E = -5.7$ mV nm$^{-1}$, measured at 15 mK.

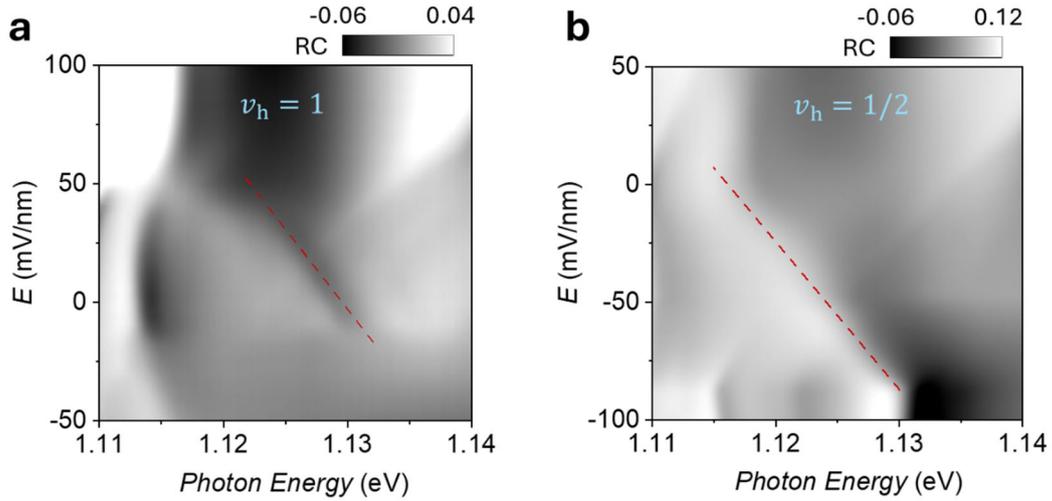

**Extended Data Figure 5 | Energy dispersion of interlayer exciton with electric field.**

**a,b,** Reflectance contrast spectrum as a function of the applied out-of-plane electric field at representative filling factors $v_\mathrm{h} = 1$ (**a**), and $1/2$ (**b**). The exciton states denoted by the dashed lines show dispersions almost linearly with $E$-field, which is a typical behavior of the Stark shift of interlayer exciton transitions, as also reported in earlier work[35]. Effective interlayer electric dipole moment $D_e \approx 0.16\, e \cdot \mathrm{nm}$ is extracted based on $\Delta E_\mathrm{iX} = D_e \Delta E$ at both filling factors (see Methods).

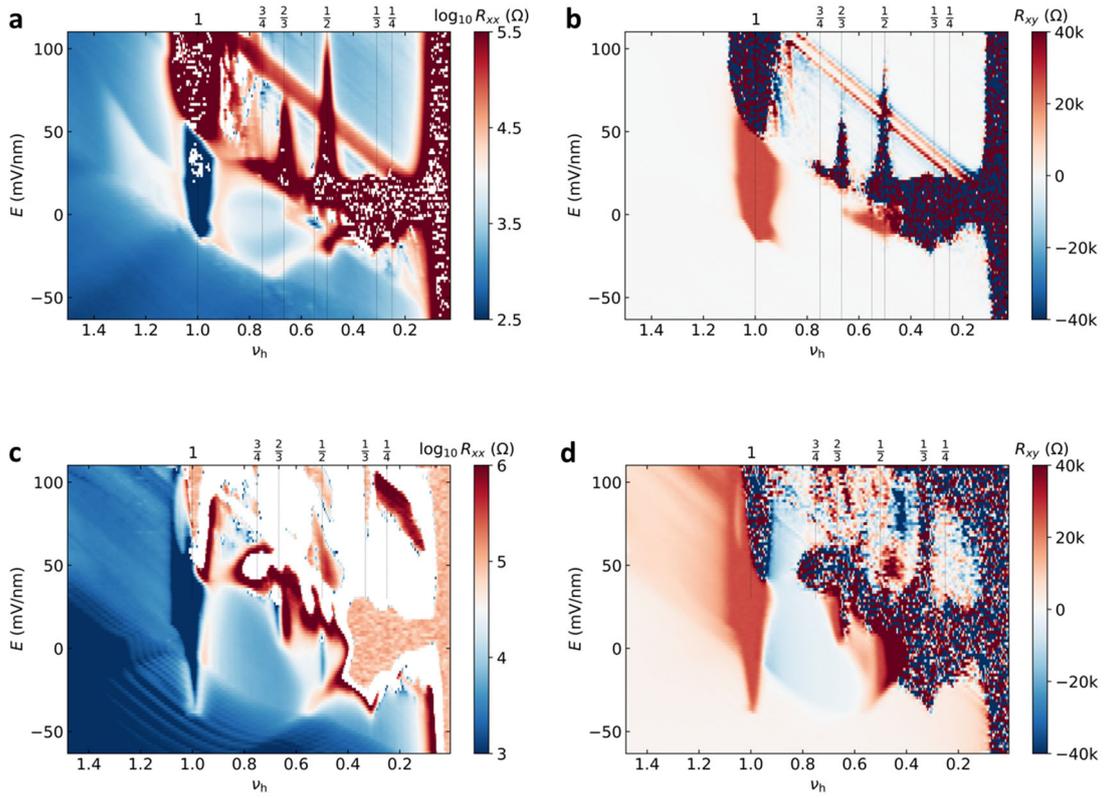

**Extended Data Figure 6 | $R_{xx}$ and $R_{xy}$ map at $B = 0.1$ and $5$ T.**

**a,b,** $R_{xx}$ and $R_{xy}$ as a function of $E$ and $\nu_h$, symmetrized (anti-symmetrized) at $B = \pm 0.1$ T, respectively. **c,d,** $R_{xx}$ and $R_{xy}$ as a function of $E$ and $\nu_h$, symmetrized (anti-symmetrized) at $B = \pm 5$ T, respectively. The $\nu_h \sim 0.53$ state vanishes at this $B$-field. $\nu_h = 1, 3/4, 2/3, 1/2, 1/3$ and $1/4$ are indicated by the vertical dashed lines. Measurements are performed at $15$ mK.

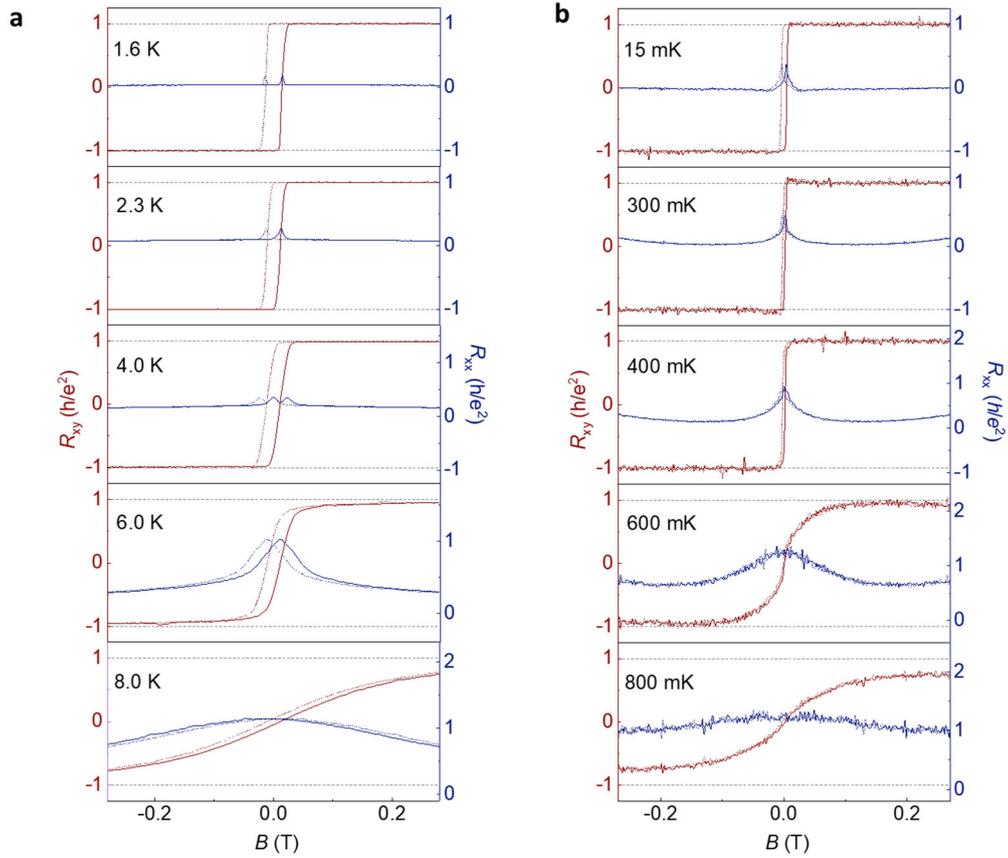

**Extended Data Figure 7 | Temperature dependence of $v_h = 1$ and $0.53$.**

**a,** $B$ dependence of $R_{xy}$ and $R_{xx}$ at representative temperatures of $v_h = 1$ at $E = -5.7$ mV nm$^{-1}$. **b,** $B$ dependence of $R_{xy}$ and $R_{xx}$ at representative temperatures of $v_h = 0.53$ at $E = -5.7$ mV nm$^{-1}$. The quantized $R_{xy}$ values are indicated by the dashed lines. $R_{xy}$ starts to deviate from the quantized value at zero magnetic field, and the hysteresis vanishes for $T$ above ~400 mK.

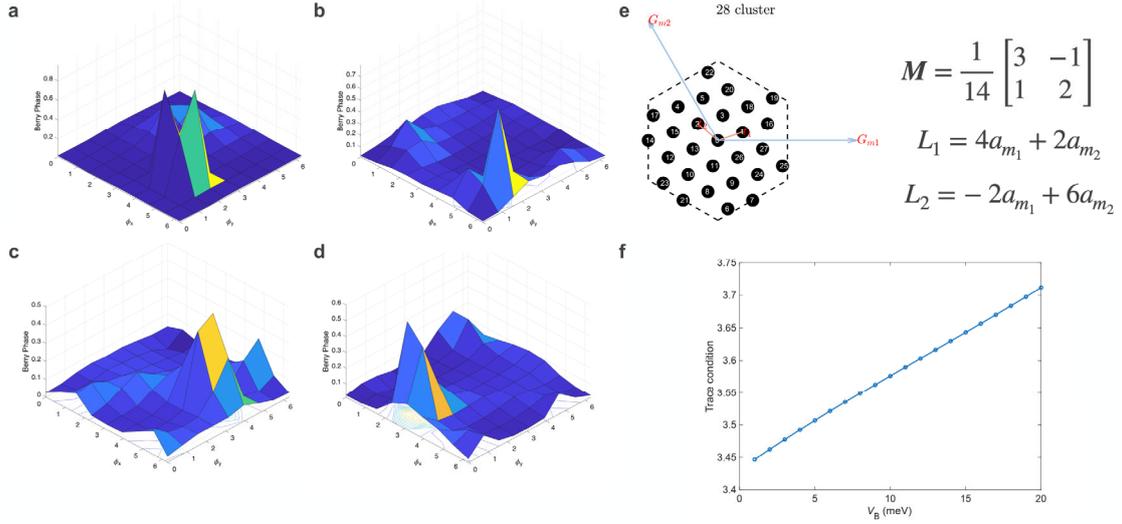

**Extended Data Figure 8 | The many-body Chern number of $v_h = 1/2$ state.**

**a,** Berry phase from each square in the boundary phase with $\phi_x \in (0, 2\pi), \phi_y \in (0, 2\pi)$, which is calculated from the many-body ground states at $k = 0$ with the parameter $\theta = 3.7°$, $V_B = 1$ meV. And we note the integral of the total Berry curvature just equals to 1. **b-d,** the same as (**a**) but at the momenta 7, 14, and 21, respectively. **e,** The finite-size cluster used in the ED calculation. $L_1$ and $L_2$ are the real-space lattice vectors of the cluster; $G_{m1}$ and $G_{m2}$ are the reciprocal vectors of the moiré system. $M$ is the matrix that relates $t_1$ and $t_2$ to $G_{m1}$ and $G_{m2}$ via $t_1 = M_{11}G_{m1} + M_{21}G_{m2}$, $t_2 = M_{12}G_{m1} + M_{22}G_{m2}$ and $k = K_1 t_1 + K_2 t_2$. **f,** Dependence of the trace condition ($T = \frac{1}{A_{BZ}} \int d^2k \, [\text{Tr}(\boldsymbol{g}(k)) - \boldsymbol{\Omega}(k)]$, $A_{BZ}$ is the area of Brillouin zone, $\boldsymbol{g}(k)$ is the quantum metric and $\boldsymbol{\Omega}(k)$ is the Berry curvature) on the strength of the intralayer potential in the B layer.

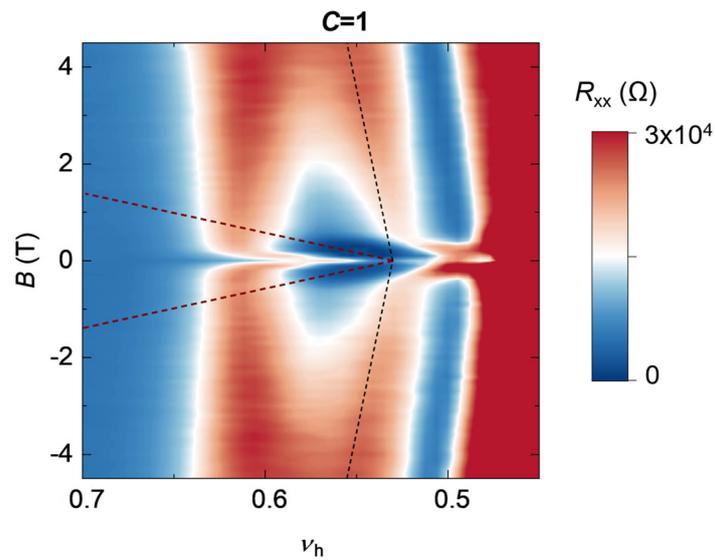

**Extended Data Figure 9 | Magnetic field dependence of the $v_h \sim 0.53$ state.** The zoomed-in view of Fig.4b. The red dashed lines represent the fitted slope using the low field data ($B < 0.5$ T). The black dashed lines represent the expected dispersions of $|C| = 1$. Above 0.5 T, the $R_{xx}$ dip is too broad to do reliable fitting.